\documentclass[11pt]{article}
\def\vec#1{\ifmmode
\mathchoice{\mbox{\boldmath$\displaystyle\bf#1$}}
{\mbox{\boldmath$\textstyle\bf#1$}}
{\mbox{\boldmath$\scriptstyle\bf#1$}}
{\mbox{\boldmath$\scriptscriptstyle\bf#1$}}\else
{\mbox{\boldmath$\bf#1$}}\fi}

\begin{document}


\begin{center}
{\Large Power-Constrained Limits}
\end{center}

\vspace{3 cm}

\begin{center}
Glen Cowan$^1$, Kyle Cranmer$^2$,
Eilam Gross$^3$, Ofer Vitells$^3$
\end{center}

\vspace{0.5 cm}

\noindent 
$^1$  Physics Department, Royal Holloway, University of London, 
Egham, TW20 0EX, U.K. \\
$^2$   Physics Department, New York University, New York, NY 10003, 
U.S.A. \\
$^3$  Weizmann Institute of Science, Rehovot 76100, Israel

\vspace{3 cm}

\begin{abstract}
We propose a method for setting limits that avoids excluding parameter
values for which the sensitivity falls below a specified threshold.
These ``power-constrained'' limits (PCL) address the issue that
motivated the widely used $\mbox{CL}_{\rm s}$ procedure \cite{cls},
but do so in a way that makes more transparent the properties of the
statistical test to which each value of the parameter is subjected.  A
case of particular interest is for upper limits on parameters that are
proportional to the cross section of a process whose existence is not
yet established.  The basic idea of the power constraint can easily be
applied, however, to other types of limits.
\end{abstract}


\clearpage

\section{Introduction}
\label{sec:intro}

In particle physics experiments one often tests specific models that
predict new phenomena.  Some regions of a model's parameter space may
be rejected by these tests; in other regions the model may be deemed
compatible with the data.  This is often done in the framework of a
frequentist statistical test, which is inverted to determine a
confidence interval.  This formalism is reviewed in
Sec.~\ref{sec:confint}.

It is generally the case that for some parameter values of a signal
model, the magnitude of the predicted effect with respect to the
background-only model is extremely small.  That is, one has
effectively no experimental sensitivity to those parts of the model's
parameter space.  Nevertheless, procedures based on frequentist tests
may exclude these values.  We discuss how this can occur and how it
has been dealt with in the past in Sections~\ref{sec:spur} and
\ref{sec:cls}.

In Sec.~\ref{sec:pcl} we introduce a new method for constraining
confidence intervals in a way that prevents one from excluding
parameter values to which one does not have sufficient sensitivity.
As the measure of sensitivity is based on the power of a statistical
test, we refer to the bounds established by these modified intervals
as power-constrained limits (PCL).

Section~\ref{sec:gausslim} illustrates the procedure for the case of
an upper limit derived from a Gaussian measurement.
Section~\ref{sec:nuisance} discusses how the procedure can be applied
in cases where there are additional nuisance parameters, beyond the
parameters of interest, that must be fitted using the data.  A summary
and conclusions are given in Sec.~\ref{sec:conclusions}.

\section{Confidence intervals from inverting a statistical test}
\label{sec:confint}

In this section we review the formalism of inverting a frequentist
statistical test to obtain a confidence interval.  A more thorough
treatment can be found in many texts, such as Ref.~\cite{Kendall2}.

We consider a test for a parameter $\mu$, which here represents the
signal strength (or any parameter proportional to the rate) of a
certain signal process.  A test of a given $\mu$ is carried out by
specifying a region of data outcomes called the {\it critical region},
which are disfavoured, in a sense discussed below, under assumption of
$\mu$.  The data outcome could be, for example, the number of events
observed in a given region of phase space, or it could represent a
larger set of numerical values.  Here we will use $\vec{x}$ to
represent the data, and $w_{\mu}$ to denote the critical region.

The critical region is chosen to such that the probability to observe
the data in it, under assumption of the hypothesized $\mu$, is not
greater than a given constant $\alpha$, called the {\it size} or {\it
significance level} of the test, i.e.,

\begin{equation}
\label{eq:sizeoftest}
P(\vec{x} \in w_{\mu} | \mu) \le \alpha \;.
\end{equation}

\noindent Often by convention $\alpha = 0.05$ is used.  If the data
are observed in the critical region, the hypothesis $\mu$ is rejected.
It is necessary in general to specify Eq.~(\ref{eq:sizeoftest}) as an
inequality because the data may be discrete (e.g., an integer number
of events), and so there may not exist a subset of the possible data
values for which the summed probability is exactly equal to $\alpha$.

It is convenient to construct from the data a test statistic
$q_{\mu}$, such that greater $q_{\mu}$ reflects an increasing level of
incompatibility between the data and the hypothesized parameter value
$\mu$.  In this way the boundary of the critical region in data space
is given by a surface of constant $q_{\mu}$, with the critical region
containing the data that give the greatest values of $q_{\mu}$.  Once
such a function has been defined, one can for any observed value
$q_{{\mu},\rm obs}$ compute a $p$-value, i.e., the probability under
assumption of $\mu$ to find data with equal or greater incompatibility
with $\mu$,

\begin{equation}
\label{eq:pval}
p_{\mu} = \int_{q_{{\mu},\rm obs}}^{\infty} f(q_{\mu} | \mu) 
\, d q_{\mu} \;,
\end{equation}

\noindent where $f(q_{\mu} | \mu)$ represents the probability density
function (pdf) of $q_{\mu}$ assuming a data distribution with strength
parameter $\mu$.  Thus the test can be equivalently formulated by
rejecting $\mu$ if its $p$-value is found less than $\alpha$.

A test of size $\alpha$ can be carried out for all values of $\mu$.
The set of values not rejected constitutes a {\it confidence interval}
for $\mu$ with confidence level $1 - \alpha$.  This interval will by
construction include the true value of the parameter with a
probability of at least $1 - \alpha$.

The procedure described above for constructing a confidence interval
by inverting a test is not unique, however, because there are (often
infinitely) many different subsets of the data space that could be
chosen for the test's critical region $w_{\mu}$.  This is usually
selected such that the probability to find $\vec{x} \in w_{\mu}$ is
large if a given alternative hypothesis (or set of alternatives) is
true.  The {\it power} of the test with respect to an alternative
value of the parameter $\mu^{\prime}$, which we denote here as
$M_{\mu^{\prime}}(\mu)$, is

\begin{equation}
\label{eq:powerdef}
M_{\mu^{\prime}}(\mu) = P(\vec{x} \in w_{\mu} | \mu^{\prime}) \;.
\end{equation}

\noindent If the test of $\mu$ is formulated using a $p$-value, such
that finding $p_{\mu} < \alpha$ is equivalent to finding $\vec{x} \in
w_{\mu}$, then the power can be written equivalently as

\begin{equation}
\label{eq:powerdef2}
M_{\mu^{\prime}}(\mu) = P(p_{\mu} < \alpha | \mu^{\prime}) \;.
\end{equation}

Often the power with respect to certain alternatives is used as the
criterion according to which one chooses the critical region of a
test.  Confidence intervals obtained from inverting the test thus
depend on this choice.  For the present discussion, however, we will
assume that the test has been defined, and the power will be used only
to modify the resulting confidence interval so that it does not
exclude parameter values to which one does not have sufficient
sensitivity.  This concept is defined more quantitatively in the
following section.

\section{Spurious exclusion}
\label{sec:spur}

When testing a hypothesized strength parameter $\mu$, it may be that
the magnitude of the signal implied by $\mu$ is extremely small --- so
small, that the probabilities for the data are very close to what they
would be in the absence of the signal process, i.e., $\mu = 0$.  In
such a case one has little or no sensitivity to the given value of
$\mu$.

For example, Fig.~\ref{fig:nosens} illustrates a situation where there
is only a very small level of sensitivity to a given strength
parameter $\mu$.  The plot shows the pdfs of the statistic $q_{\mu}$
under assumption of strength parameters $\mu$, and also assuming $\mu
= 0$, i.e., $f(q_{\mu} | \mu)$ and $f(q_{\mu} | 0)$.  If the observed
value of the statistic is found in the critical region corresponding
to the top 5\% of $f(q_{\mu} | \mu)$, then the hypothesized $\mu$ is
rejected.  But as the two pdfs almost coincide, the probability to
reject $\mu$ if the true strength parameter is zero is also close to
$\alpha = 0.05$.

\setlength{\unitlength}{1.0 cm}
\renewcommand{\baselinestretch}{0.9}
\begin{figure}[htbp]
\begin{picture}(10.0,6.)
\put(0.6,-0.5)
{\includegraphics{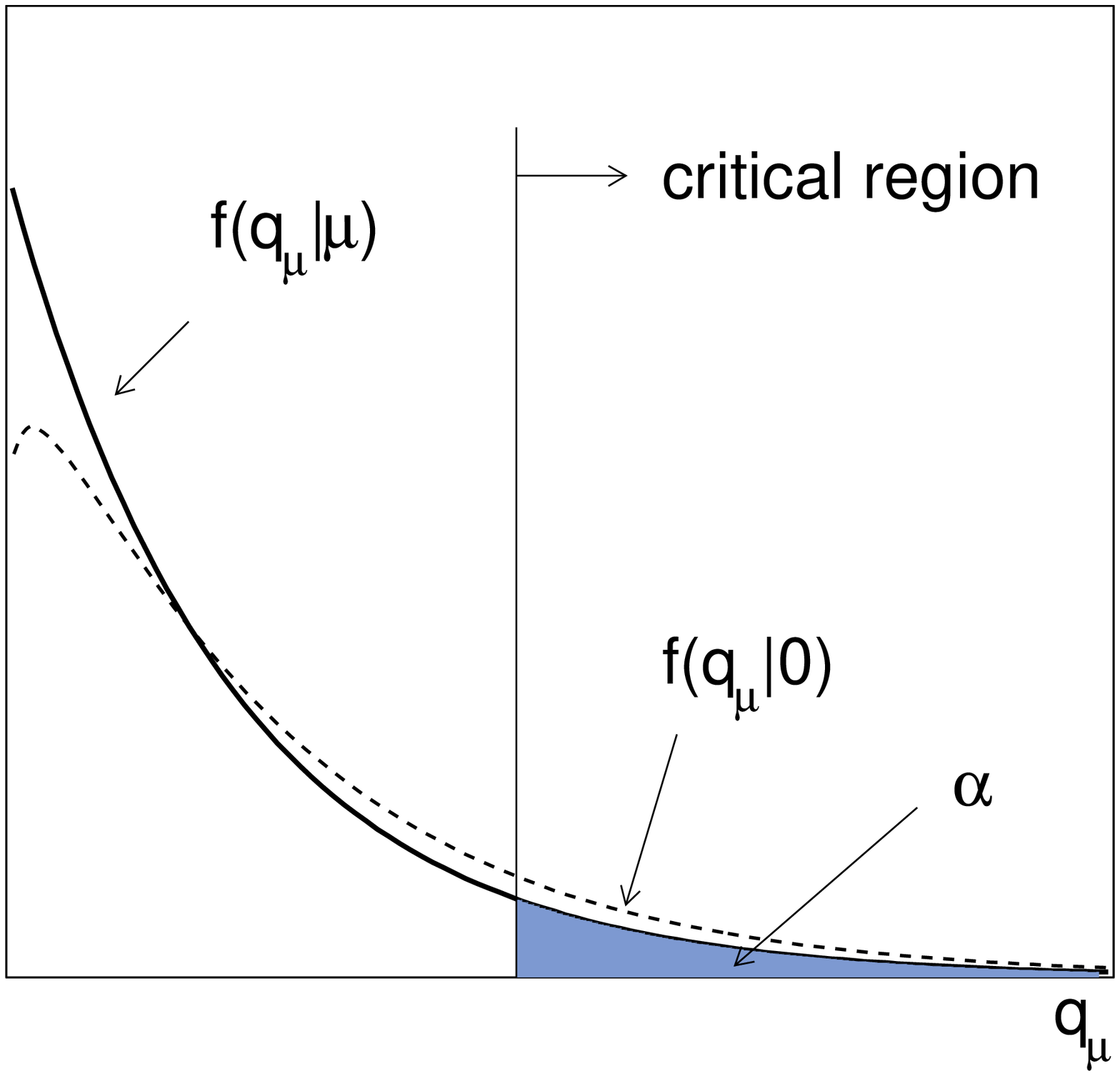}}
\put(8.1,-0.5)
{\includegraphics{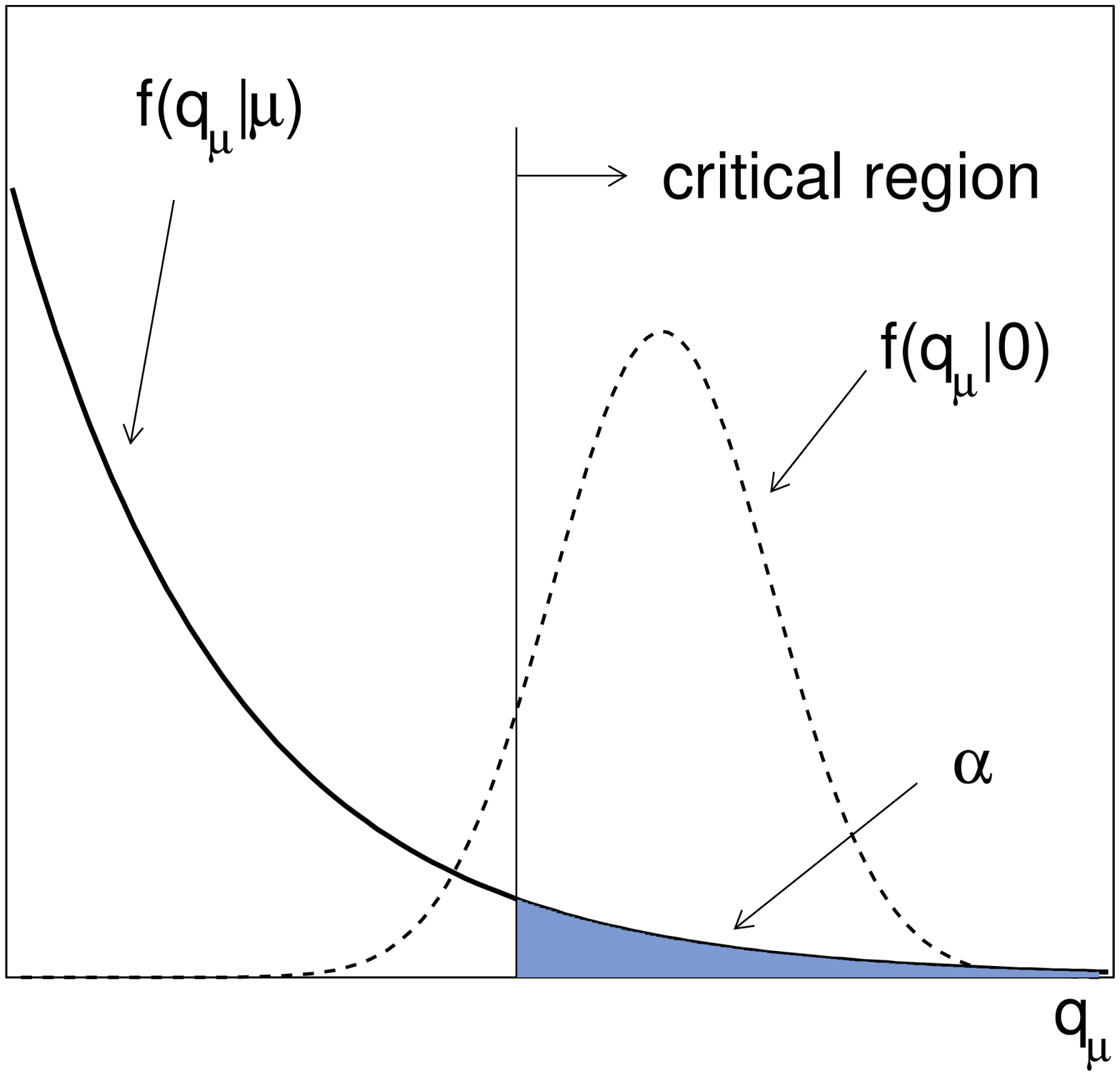}}
\put(0.1,5.){(a)}
\put(14.8,5.){(b)}
\end{picture}
\caption{\small Illustration of statistical tests of
parameter values $\mu$ for the cases of (a) little sensitivity
and (b) substantial sensitivity (see text).}
\label{fig:nosens}
\end{figure}
\renewcommand{\baselinestretch}{1}
\small\normalsize

Figure~\ref{fig:nosens}(b) shows the same distributions as (a) but for
a different value of $\mu$.  The size of the test is, as in (a), equal
to $\alpha$.  Here, however, the distribution of $q_{\mu}$ under the
assumption of $\mu^{\prime} = 0$ leads to a substantially greater
probability to reject $\mu$, i.e., to find $q_{\mu}$ in the critical
region.

The sensitivity of a test of $\mu$ can be quantified using the power
of the test with respect to a stated alternative $\mu^{\prime}$, which
we will take here to be the no-signal hypothesis, $\mu^{\prime} = 0$.
In the case where the pdfs $f(q_{\mu} | \mu)$ and $f(q_{\mu} | 0)$
coincide, the probability to reject $\mu$ assuming the alternative
$\mu^{\prime}=0$ approaches the significance level of the test,
$\alpha$.

In the context of a search for a new phenomenon, this means that with
probability not less than $\alpha$ one will exclude hypotheses to
which one has little or no sensitivity, which we refer to here as
spurious exclusion.  The hypothesis might indeed be false, but if it
is excluded, this is more naturally interpreted as a data fluctuation
away from the region favoured under assumption of $\mu$.  This could
result, for example, in a search for a hypothetical particle with a
mass far above the range where it would have a noticeable impact on
the data.  Particle Physics experiments often carry out many searches
covering a broad parameter range for many signal models, and so
spurious exclusion is in fact a problem that can arise often.

\section{Previous methods that address spurious exclusion}
\label{sec:cls}

The problem of spurious exclusion, or equivalently, having a ``lucky''
statistical fluctuation lead to an anomalously strong limit, has been
known in the particle physics community for many years.  The note by
Highland \cite{highland} reviews the problem and proposes several
possible solutions; further discussion can be found in the review on
statistics by the Particle Data Group \cite{PDG}.

The problem received particular focus during searches for the Higgs
Boson at the LEP Collider in the 1990s, and led to a procedure called
``$\mbox{CL}_{\rm s}$'' \cite{cls}.  Here one forms the ratio

\begin{equation}
\label{eq:cls}
\mbox{CL}_s = \frac{p_{\mu}}{1 - p_0} \;,
\end{equation}

\noindent where $p_{\mu}$ and $p_0$ are the $p$-values of the
hypothesized strength parameter values $\mu$ and $0$, respectively.
In the $\mbox{CL}_{\rm s}$ procedure, $\mu$ is deemed to be excluded
if one finds $\mbox{CL}_s < \alpha$.  Because $\mbox{CL}_s$ is aways
greater than $p_{\mu}$, the probability of exclusion assuming $\mu$ is
necessarily less than $\alpha$.  Thus the quoted upper limit from the
$\mbox{CL}_{\rm s}$ procedure will be greater than the upper limit
according to the method of Sec.~\ref{sec:confint}, and in this sense
the $\mbox{CL}_{\rm s}$ procedure is said to be conservative.  This is
illustrated in the example described in Sec.~\ref{sec:gausslim}.

Because of this conservatism, the frequentist coverage probability of
the $\mbox{CL}_{\rm s}$ upper limits (i.e., the probability under
assumption of $\mu$ that the interval will contain $\mu$) is not equal
to $\alpha$, but is in general larger.  Although the exact coverage
probabilities of $\mbox{CL}_{\rm s}$ intervals can be found as a
function of $\mu$, this is not often reported.

\section{Power Constrained Limits}
\label{sec:pcl}

Here we propose an alternate procedure for producing intervals whose
coverage properties are easily apparent for all values of $\mu$.  To
do this we break the range of $\mu$ to be tested into two categories
based on the power $M_{0}(\mu)$ of a test of $\mu$ with respect to the
no-signal alternative, $\mu^{\prime}=0$.  If this power is below a
specified threshold $M_{\rm min}$, one's sensitivity to this parameter
is deemed to be too low and the point is not regarded as testable.  If
the power is greater than or equal to the threshold, then the test of
size $\alpha$ is carried out.  A value of $\mu$ is excluded if

\begin{enumerate}
\item[(a)] the value $\mu$ is rejected by the test, i.e., 
$\vec{x} \in w_{\mu}$ or equivalently $p_{\mu} < \alpha$, and
\item[(b)] one has sufficient sensitivity to $\mu$, i.e., 
$M_{0}(\mu) \ge M_{\rm min}$.
\end{enumerate}

\noindent An interval is constructed from the values of $\mu$ not
excluded.  If this is done on the basis of the test (a) only, it is
referred to here as an {\it unconstrained} interval.  Application of
the {\it power constraint} (b) results in the power-constrained
interval or limit.

The coverage probability of the power-constrained interval is $100\%$
for $\mu$ values that have power below $M_{\rm min}$, and $\alpha$ for
those values with power greater than or equal to the threshold.  When
reporting the result it is recommended to indicate which parameter
values were above and which below the power-constraint threshold, and
in this way one can easily see what the coverage probability is for
all values of $\mu$.

The choice of the minimum power threshold is a matter of convention.
We prefer to use $M_{\rm min} = 0.16$, or more precisely, $M_{\rm min}
= \Phi(-1) = 0.1587$, where $\Phi$ is the standard normal cumulative
distribution (i.e., the cumulative distribution for Gaussian with a
mean of zero and unit standard deviation).  As shown below, this
corresponds to applying the power constraint if the unconstrained
limit fluctuates one standard deviation below its median value under
the background-only hypothesis.

This procedure bears some similarity to one introduced recently in the
astrophysics community in Ref.~\cite{astropcl}, although there the
power refers to a test of the background-only ($\mu=0$) hypothesis,
and furthermore the result is not used in quite the same way as what
we propose here.  Note also in Ref.~\cite{astropcl}, ``upper bound''
is similar to what we call an upper limit, and their term ``upper
limit'' is taken to refer to the sensitivity threshold.

Formally, to construct the interval for $\mu$ one begins by finding
the power for a test of each $\mu$ with respect to the alternative
$\mu^{\prime}=0$,

\begin{equation}
\label{eq:maechtigkeit}
M_{0}(\mu) = P(\vec{x} \in w_{\mu}|0) = P( p_{\mu} < \alpha | 0 ) \;.
\end{equation}

\noindent In some problems this can be found in closed form; otherwise
it can be obtained using a Monte Carlo calculation, in which one for
every value of $\mu$ calculates the distribution of $p_{\mu}$ using
data generated according to $\mu=0$.  The value $M_{0}(\mu)$ is then
found simply by integrating each distribution from zero up to the
desired significance level $\alpha$ (e.g., $0.05$).  

An equivalent and in ways simpler procedure is first to carry out the
statistical test without the power constraint, and invert this to find
the unconstrained confidence interval for $\mu$.  Some of the parameter
values that are excluded from this interval may be found to have
a power below the required threshold, and they are then re-included
in the power-constrained interval, which is thus by construction 
a superset of the unconstrained one.

For example, one may be interested in finding an an upper limit,
$\mu_{\rm up}$, i.e., the largest value of $\mu$ not excluded.  By
inverting the test, one determines $\mu_{\rm up}$ as a function of the
data.  One can therefore determine the distribution of $\mu_{\rm up}$,
e.g., by simulating the experiment many times under assumption of
$\mu=0$ and constructing a histogram of $\mu_{\rm up}$ for each
outcome.  Then for each value of $\mu$ one determines the
corresponding power.  This is the probability, under assumption of the
background-only ($\mu=0$) hypothesis, to reject $\mu$, i.e., to find
$\mu$ outside of the unconstrained confidence interval.  In the case
of an upper limit this is

\begin{equation}
\label{eq:powerfrommuup}
M_0(\mu) = P(\mu_{\rm up} < \mu | 0 ) \;.
\end{equation}

One should note the following caveat: It can be that for certain data
outcomes, all values of $\mu$ are excluded by the test, in which case
$\mu_{\rm up}$ is not defined.  In such cases one must count the
outcomes as contributing to the probability that $\mu$ is outside the
confidence interval.

With this in mind, one can then find the smallest value of $\mu$ for
which the power $M_0(\mu)$ is at least equal to the minimum value
$M_{\rm min}$, denoted here as $\mu_{\rm min}$.  The Power-Constrained
Limit $\mu^{*}_{\rm up}$ is given by the larger of the unconstrained
limit $\mu_{\rm up}$ or the minimum value to which one has
sensitivity, $\mu_{\rm min}$:

\begin{equation}
\label{eq:pclmuup}
\mu^{*}_{\rm up} = \mbox{max}(\mu_{\rm up}, \mu_{\rm min}) \;.
\end{equation}

\section{PCL for an upper limit based on a Gaussian measurement}
\label{sec:gausslim}

Often the test of $\mu$ is based on a Gaussian distributed
measurement.  For example, for a sufficiently large data sample and
under conditions often satisfied in practice, the distribution of the
Maximum Likelihood Estimator $\hat{\mu}$ has a Gaussian form with
standard deviation $\sigma$ and is centred about the true $\mu$.  Here
we will assume this is the case and further take $\sigma$ to be known.


For the case of an upper limit, we define the critical region to
contain the lowest values of $\hat{\mu}$ such that the probability to
find $\hat{\mu}$ there is equal to $\alpha$.  For Gaussian distributed
$\hat{\mu}$ with mean $\mu$ and standard deviation $\sigma$,
the critical region is therefore

\begin{equation}
\label{eq:muhatc}
\hat{\mu} < \mu - \sigma \Phi^{-1}(1 - \alpha) \;,
\end{equation}

\noindent where $\Phi^{-1}$ is the inverse of the standard Gaussian
cumulative distribution (the standard normal quantile).  For example,
$\alpha = 0.05$ gives $\Phi^{-1}(1-\alpha) = 1.64$.

Rejecting $\mu$ if the data are in the critical region gives the
unconstrained upper limit,

\begin{equation}
\label{eq:muup1}
\mu_{\rm up} = \hat{\mu} + \sigma \Phi^{-1}(1 - \alpha) \;.
\end{equation}

The power of the test of $\mu$ with respect to the alternative
$\mu^{\prime} = 0$ is

\begin{equation}
\label{eq:M0mu}
M_{0}(\mu) = P \left(
\hat{\mu} < \mu - \sigma \Phi^{-1}(1 - \alpha) | 0 \right) \;.
\end{equation}

\noindent Because $\hat{\mu}$ here follows a Gaussian distribution,
the power can be written

\begin{equation}
\label{eq:M0mu2}
M_{0}(\mu) = \Phi \left( \frac{\mu}{\sigma} - \Phi^{-1}(1 - \alpha) \right)
\;.
\end{equation}

\noindent This is illustrated in Fig.~\ref{fig:powerfunc} for $\alpha
= 0.05$ and $\sigma = 1$.  Since the cumulative distribution $\Phi$ is
monotonically increasing and furthermore $\Phi(1 - \alpha) = -
\Phi(\alpha)$, Eq.~(\ref{eq:M0mu2}) gives $M_{0}(0) = \alpha$ and
$M_{0}(\mu) > \alpha$ for all $\mu > 0$, as can be seen in the figure.

\setlength{\unitlength}{1.0 cm}
\renewcommand{\baselinestretch}{0.8}
\begin{figure}[htbp]
\begin{picture}(10.0,6)
\put(1,0.){\includegraphics{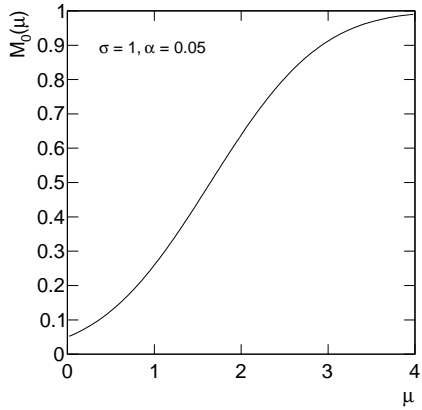}}
\put(9.0,0.){\makebox(5,4)[b]{\begin{minipage}[b]{5cm}
\protect\caption{{\footnotesize The power $M_{0}(\mu)$ for a
test of $\mu$ with respect to the alternative $\mu^{\prime} = 0$ (see
text).}  
\protect\label{fig:powerfunc}}
\end{minipage}}}
\end{picture}
\end{figure}
\renewcommand{\baselinestretch}{1}
\small\normalsize

Requiring the power $M_{0}(\mu) \ge M_{\rm min}$,

\begin{equation}
\label{eq:powercon}
\Phi \left( \frac{\mu}{\sigma} - \Phi^{-1}(1 - \alpha) \right)
\ge M_{\rm min} \;,
\end{equation}

\noindent implies that the smallest $\mu$ to which one is
sensitive is

\begin{equation}
\label{eq:mumin}
\mu_{\rm min} = \sigma \left( 
\Phi^{-1}(M_{\rm min}) + \Phi^{-1}(1 - \alpha)  \right) \;.
\end{equation}

By combining Eqs.~(\ref{eq:muup1}) and (\ref{eq:mumin}), one
sees that $\mu_{\rm up}$ is below $\mu_{\min}$ if one finds

\begin{equation}
\label{eq:muhatcon1}
\hat{\mu} < \sigma \Phi^{-1}(M_{\rm min}) \;.
\end{equation}

\noindent Thus one finds the following expression for the
power-constrained upper limit:

\begin{equation}
\label{eq:muup}
\mu^{*}_{\rm up} = \left\{ \! \! 
               \begin{array}{ll}
               \sigma \left( \Phi^{-1}(M_{\rm min}) + 
                             \Phi^{-1}(1 - \alpha) \right)  & \quad \quad
               \hat{\mu} < \sigma \Phi^{-1}(M_{\rm min}) \;, \\*[0.3 cm]
               \hat{\mu} + \sigma \Phi^{-1}(1 - \alpha) & 
               \quad \quad \mbox{otherwise} \;.
               \end{array} 
               \right. 
\end{equation}

\noindent This is shown as a function of $\hat{\mu}$ in
Fig.~\ref{fig:muup}(a).

\setlength{\unitlength}{1.0 cm}
\renewcommand{\baselinestretch}{0.9}
\begin{figure}[htbp]
\begin{picture}(10.0,6.)
\put(0.5,-0.5)
{\includegraphics{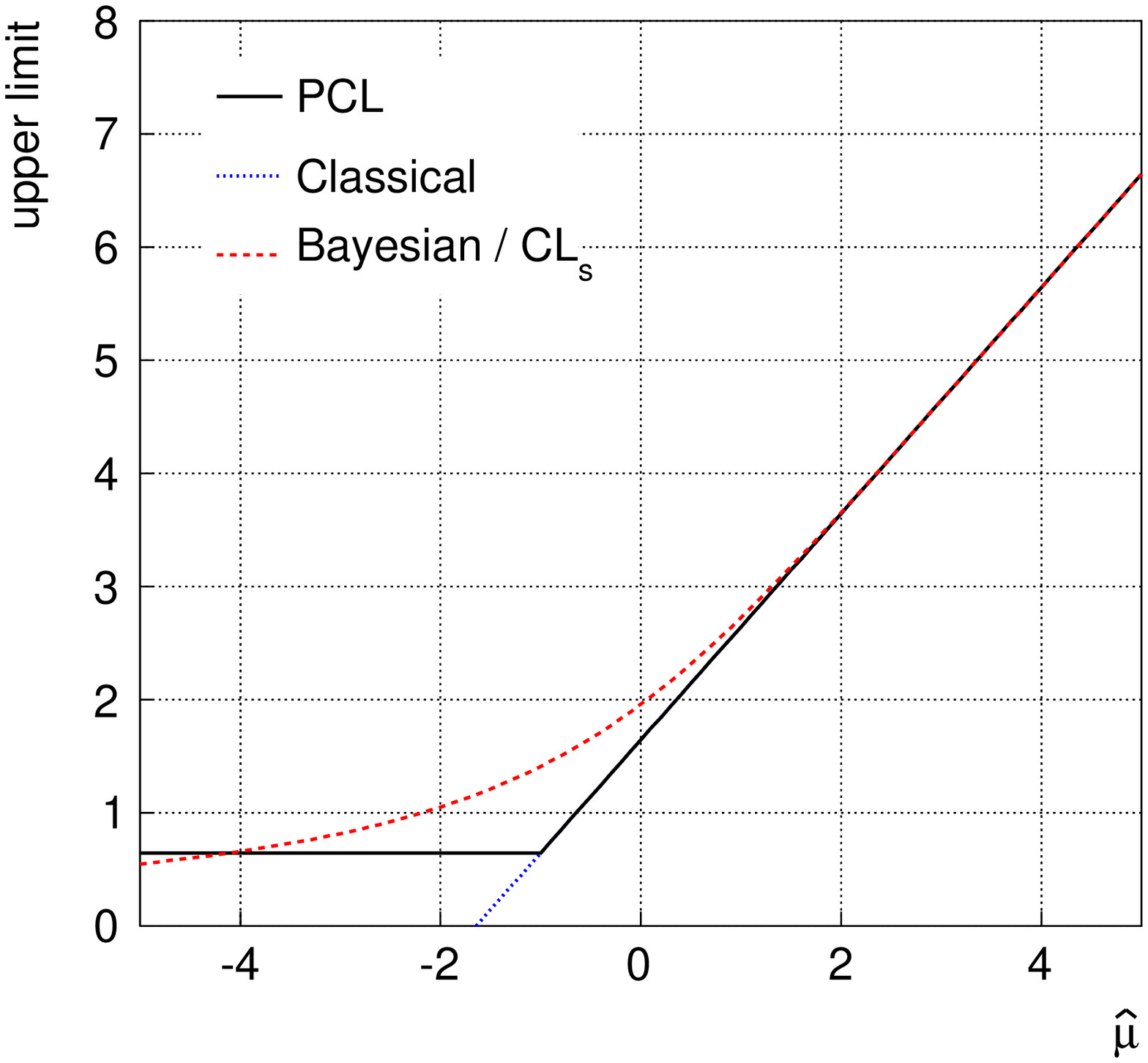}}
\put(8,-0.5)
{\includegraphics{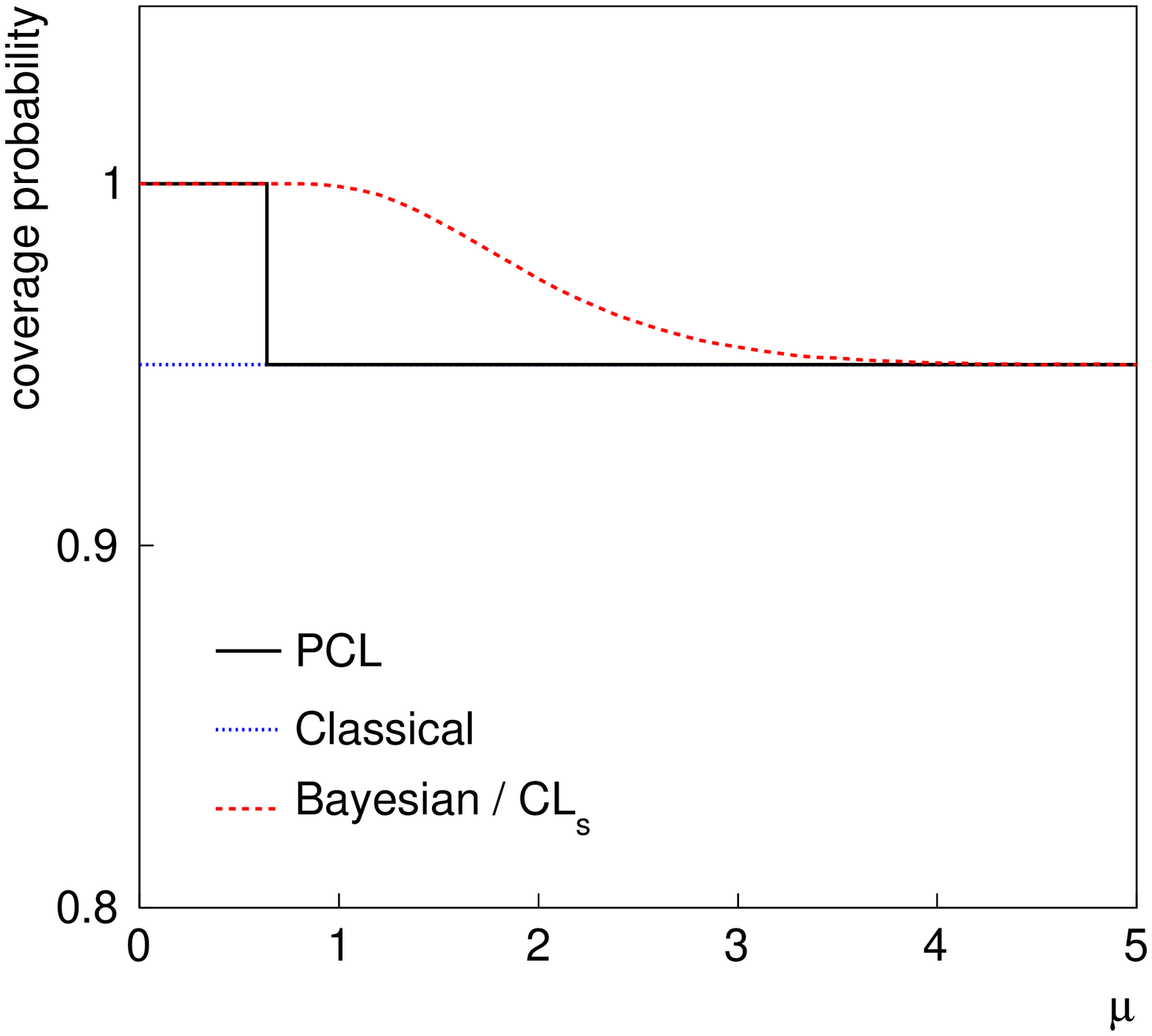}}
\put(0.1,5.5){(a)}
\put(14.9,5.5){(b)}
\end{picture}
\caption{\small (a) Upper limits from the PCL (solid),
$\mbox{CL}_{\rm s}$ and Bayesian (dashed), and classical (dotted)
procedures as a function of $\hat{\mu)}$,
which is assumed to follow a Gaussian distribution with
unit standard deviation. (b) The corresponding 
coverage probabilities as a function of $\mu$.}
\label{fig:muup}
\end{figure}
\renewcommand{\baselinestretch}{1}
\small\normalsize

For comparison, Fig.~\ref{fig:muup}(a) also shows the upper limit
without the power costraint (here called ``classical'') as well as the
one obtained from the $\mbox{CL}_{\rm s}$ procedure, which for this
particular problem coincides with the Bayesian upper limit when using
a constant prior for $\mu \ge 0$.

Figure~\ref{fig:muup}(b) shows the corresponding coverage
probabilities for the upper limits.  For PCL, this is 100\% for $\mu <
\mu_{\min} = \sigma (\Phi^{-1}(M_{\rm min}) + \Phi^{-1}(1-\alpha)) =
0.64$, and 95\% otherwise.  For $\mbox{CL}_{\rm s}$ and Bayesian, the
coverage probability is everywhere greater than 95\%, approaching 95\%
as $\mu$ increases.

\section{Distribution of upper limit and choice of minimum power}
\label{sec:minpow}

As mentioned above, we prefer to take the minimum power threshold
$M_{\rm min} = \Phi(-1) = 0.1587$.  From Eq.~(\ref{eq:muhatcon1}) one
can see that if $\mu_{\rm up}$ follows a Gaussian distribution, this
choice of $M_{\rm min}$ corresponds to applying the power constraint
if the data fluctuate below their expected value, under assumption of
$\mu=0$, by more than one standard deviation.  Here we will refer to a
fluctuation at this level as $1 \sigma$ (downward), regardless of the
distribution of $\mu_{\rm up}$.  In fact, the distribution of
$\mu_{\rm up}$ often is close to Gaussian so the terminology is
natural and convenient.

This choice of $M_{\rm min}$ can be motivated by the idea that a
sufficiently small fluctuation should not result in spurious exclusion
of the type that the PCL and $\mbox{CL}_{\rm s}$ procedures are
intended to prevent.  If, for example, one were to require $M_{\rm
min} = 0.5$, then one would impose the power constraint whenever the
observed limit is found below the median, i.e., half of the time,
which is not consistent with the notion of accepting small
fluctuations.  Therefore we feel requiring a power of 50\% is too
extreme.

On the other hand, for any (unbiased) test, the power is always
greater than or equal to the significance level $\alpha$.  So if one
were to take $M_{\rm min} \le \alpha$ then the result is the same as the
unconstrained limit.  Since one often takes $\alpha = 0.05$, taking
$M_{\rm min} = 0.05$ would correspond to a $1.64 \sigma$ downward
fluctuation (i.e., $\Phi(-1.64) = 0.05$).  

Sensitivity to the parameter $\mu$ corresponds having a power
$M_{0}(\mu)$ substantially larger than the significance level
$\alpha$.  Therefore one would like to take $M_{\rm min}$ large with
respect to $\alpha$, while still allowing for moderate a downward
fluction of the limit before imposing the power constraint.  We
therefore believe $M_{\rm min} = \Phi(-1) \approx 0.16$ is a natural
choice for use with $\alpha = 0.05$.  This allows for fluctuations up
to the one-sigma level before imposing the power constraint, and the
difference between $\alpha = 0.05$ and $M_{\rm min} = 0.16$ is
sufficient to ensure a reasonable sensitivity.  If one were to take,
e.g., $\alpha = 0.1$, as is done in some analyses, then one may
consider that a somewhat larger $M_{\rm min}$ is appropriate.

In many searches for new phenomena, one may carry out the analysis for
a range of parameters in the signal model.  For example, when
searching for the Higgs boson one may carry out the analysis for each
value of the mass $m_{\rm H}$.  In this situation one can simply
repeat the power-constraint procedure for each value of the signal
model's parameters, as is illustrated in Fig.~\ref{fig:pclband}.

\setlength{\unitlength}{1.0 cm}
\renewcommand{\baselinestretch}{0.8}
\begin{figure}[htbp]
\begin{picture}(10.0,5.5)
\put(1.0,0.2){\includegraphics{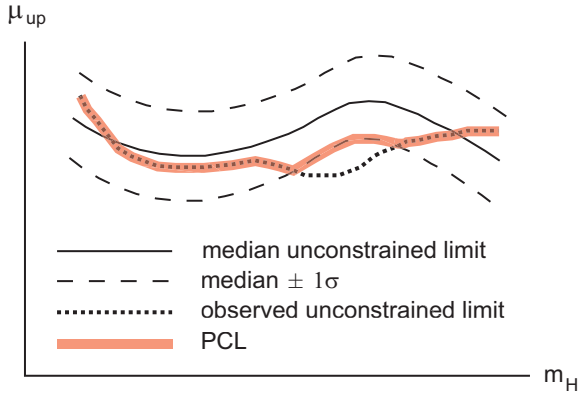}}
\put(10.0,0.){\makebox(5,4)[b]{\begin{minipage}[b]{5cm}
\protect\caption{{\footnotesize Illustration of the power-constrained
limit as a function of a model parameter such as the Higgs boson
mass $m_{\rm H}$ (see text).}  
\protect\label{fig:pclband}}
\end{minipage}}}
\end{picture}
\end{figure}
\renewcommand{\baselinestretch}{1}
\small\normalsize

In Fig.~\ref{fig:pclband}, the solid line represents the median value
of the unconstrained upper limit $\mu_{\rm up}$, and the lower and
upper dashed curves are the 0.16 and 0.84 quantiles of the
distribution of $\mu_{\rm up}$.  The dotted curve in
Fig.~\ref{fig:pclband} represents a possible outcome for the
unconstrained limit $\mu_{\rm up}$.  The minimum power is taken to be
$M_{\rm min} = \Phi(-1) = 0.16$, and thus the power-constrained limit
is the greater of the dotted and lower dashed curves, as indicated
by the shaded curve.

\section{Treatment of nuisance parameters}
\label{sec:nuisance}

In many analyses, the probability model that describes the data is not
uniquely specified by the parameter (or parameters) of interest, but
rather also contains nuisance parameters.  That is, the values of
these parameters are not known a priori and they must be fitted using
the data.  For concreteness suppose the model is characterized by a
strength parameter $\mu$ and a set of nuisance parameters
$\vec{\theta} = (\theta_1, \ldots, \theta_N)$.

The nuisance parameters complicate the present problem in two ways.
First, they make it difficult to construct an unconstrained interval
for the parameter of interest that has the correct coverage
probability for all values of $\vec{\theta}$.  This problem has been
widely discussed in recent years, e.g., Ref.~\cite{phystat05}.  Many
of the proposed procedures give intervals with correct coverage for
some values of $\vec{\theta}$, but approximate coverage elsewhere.
For example, an approximate solution based on the profile likelihood
ratio test is discussed in Refs.~\cite{asimov}.  For the present
discussion we will assume that a test procedure that gives an
unconstrained interval has been chosen.  Its coverage probability may
or may not be exactly equal to the nominal confidence level for all
values of $\vec{\theta}$.

Of more direct concern for the present paper is the fact that the
power of the test of $\mu$ with respect to the no-signal alternative
will depend in general on the nuisance parameters $\vec{\theta}$.  As
the power is intended to represent the probability, under assumption
of the no-signal model, to reject a given value of $\mu$, we take the
values of $\vec{\theta}$ that are in best agreement with the actual
data under assumption of $\mu=0$.  We denote these as
$\hat{\hat{\vec{\theta}}}(0)$, i.e., they are the conditional
estimators for $\vec{\theta}$ under assumption of $\mu=0$.

As a consequence of this choice, the power $M_0(\mu)$ becomes a
function of the actual data, since the data are used to determine
values for the nuisance parameters.  Thus the range of $\mu$ values
where one has sufficient sensitivity also depends to some extent on
the data.  This may seem counter-intuitive, since the power of a
specific test, i.e., at a given point in $(\mu, \vec{\theta})$-space,
is independent of the data.  But there is a certain power $M_0(\mu)$
for every point in $\vec{\theta}$-space, and one uses the data to
choose the point at which one quotes the power.

Alternatively, one could require that the power is greater than or
equal to the minimum threshold for all values of the nuisance
parameters in a specified range.  In this way the set of $\mu$
values for which one has sufficient sensitivity would not depend on
the data.  As this would entail considerable computational effort,
however, we prefer to define the power using a specific point in the
nuisance-parameter space as described above.

\section{Summary and conclusions}
\label{sec:conclusions}

We propose a power-constraint procedure for modifying confidence
limits so that parameter values to which one has little or no
sensitivity are not excluded.  The sensitivity is measured using the
power of the test of the parameter with respect to the no-signal
alternative.  The coverage probability of the resulting limits is
equal to the nominal confidence level (e.g., 95\%) for parameter
values to which one's sensitivity is above a given threshold, and
100\% if the sensitivity is below the threshold.  This can be
contrasted with the $\mbox{CL}_{\rm s}$ procedure, for which the
coverage probability is always greater than the nominal confidence
level by an amount that varies continuously as a function of the
assumed parameter value.

The power used for the sensitivity threshold is a matter of
convention, but we recommend taking this to be $M_{\rm min} = \Phi(-1)
\approx 0.16$.  This is consistent with allowing for reasonably small
downward fluctuations of the data by drawing the boundary at the
one-sigma level.  Allowing more than $1.64 \sigma$ fluctuations would
mean the power constraint is never imposed (for a 95\% confidence
level limit), and requiring $M_{\rm min} = 0.5$ would impose the power
constraint half of the time, including cases with only an
infinitesimal downward fluctuation.

The PCL procedure is easily extended to problems with nuisance
parameters.  There we define the power with respect to the
background-only ($\mu=0$) model using the conditional estimates of the
nuisance parameters given $\mu=0$.

The PCL procedure is particularly useful in cases where spurious
exclusion is problematic, such as when a one-sided test is inverted to
give an upper limit.  It can be applied, however, to any confidence
interval, including those based on inversion of a likelihood-ratio
test (i.e., Feldman-Cousins intervals \cite{FC}).

When reporting results, we recommend to show both the constrained and
unconstrained limits.  In this way one can know whether a given
parameter value is not rejected because the data are in good agreement
with it, or rather because it is a value to which the sensitivity is
deemed to low to allow exclusion.


\end{document}